\documentclass{calor2002}
\usepackage{epsfig}
\begin{document}

\title{Performance of the OPAL Si-W luminometer at LEP I-II}

\author{G. ABBIENDI}

\address{Dipartimento di Fisica dell'Universit\`a di Bologna and INFN,\\
Viale Berti Pichat 6/2, 40127 Bologna, Italy \\ 
E-mail: Giovanni.Abbiendi@bo.infn.it}

\author{R.~G. KELLOGG}

\address{Department of Physics, University of Maryland, 
College Park, MD 20742, USA \\
E-mail: Richard.Kellogg@cern.ch}

\author{D. STROM}

\address{University of Oregon, Department of Physics, 
Eugene OR 97403, USA \\
E-mail: strom@bovine.uoregon.edu}


\maketitle

\abstracts{
A pair of compact Silicon-Tungsten calorimeters was operated in 
the OPAL experiment at LEP to measure the integrated luminosity
from detection of Bhabha 
$e^\pm$ scattered at small angles from the beam line. 
%electrons scattered at angles between 
%25 and 58 mrad from the beam line. 
%In the eight years from 1993
%to 2000 the detector worked first at the Z mass peak and then
%at center of mass energies up to 209 GeV. The fine radial and
%longitudinal segmentation (2.5 mm $\times$ 1~$X_0$) allowed the radial
%position of electron and photon showers to be measured with a
%resolution of 130-170~$\mu$m and a residual radial bias as
%small as 7~$\mu$m. Reducing the bias in the definition of the
%inner acceptance radius was the key element in obtaining an
%experimental systematic error on the integrated luminosity of
%only $3.4\times10^{-4}$. 
The performance of the detector at both LEP-I
and LEP-II is reviewed. 
%Energy resolution, sensitivity to 
%overlapping electromagnetic showers and sensitivity to minimum
%ionizing particles are discussed.
}

\section{Introduction}

The LEP $e^+e^-$ collider at CERN operated for more than a decade: in 1989-95 at
center of mass energies close to the $Z$ peak (LEP-I); in 1996-2000 at higher
energies, up to $209$~GeV (LEP-II). In the first phase a large number of $Z^0$
events were collected, of the order of $5 \times 10^6$~events per experiment.
To match the inherent precision of this data sample, the error on the
integrated luminosity had to be better
than $10^{-3}$. 
At LEP the relevant process for the luminosity
measurement is Bhabha scattering at small angle, which delivers a counting rate
higher than the $Z^0$ event rate at resonance.
The Bhabha 
angular spectrum falls like $1/\theta^3$, implying a high sensitivity
to the definition of the minimum polar angle of the acceptance. For example an
uncertainty $\delta \theta = 10$~$\mu$rad (which in our configuration
is equivalent to 25~$\mu$m in radius at the face of the detector) would give an 
unacceptable systematic error of $10^{-3}$.
Precision luminosity measurement was thus a demanding task, dictated by 
interest in measuring absolute cross sections at the $Z^0$ peak. 
In particular, cross sections were needed
to determine the Invisible Ratio $R_{inv} = \Gamma_{inv} / \Gamma_{ll}$,
% with $\Gamma_{inv} = \Gamma_Z - \Gamma_{had} - 3
%\Gamma_{ll}$. 
the ratio of the $Z$ decay width to invisible particles 
and to charged lepton pairs.
%, while $\Gamma_Z$ and $\Gamma_{had}$ are the total
%decay width and its fraction to hadrons
From $R_{inv}$ the LEP experiments
determined the number of light neutrinos to be 3 and limited possible
contributions from extra new physics like cold dark matter.
\footnote{OPAL results are
$N_\nu = 2.984 \pm 0.013$ and $\Gamma^{new}_{inv} < 3.7$~MeV at 95$\%$
confidence level.}

\section{Detector}

The OPAL Si-W luminometer consists of 2 identical cylindrical calorimeters, 
encircling the beam pipe simmetrically at $\pm 2.5$~m from the interaction 
point. A detailed description can be found in the OPAL paper
\cite{bib:lumi}.
Each calorimeter is a stack of 19 silicon layers interleaved with 18 tungsten
plates,
%, as shown in Figure \ref{fig:swdraw}. 
with a sensitive depth of
14~cm, representing 22~$X_0$. The first 14 tungsten plates are each 1~$X_0$
thick, while the last 4 are each 2~$X_0$ thick. The sensitive area
fully covers radii between 6.2 and 14.2~cm from the beam axis, giving a
total area of silicon of 1.0~m$^2$ per calorimeter. Each silicon layer is
divided into 16 overlapping wedges.
%, so that the active areas of all adjacent
%wedges are contiguous
Even and odd layers are staggered by an azimuthal
rotation of half a wedge.
Water cooling pipes run as close as possible to the readout chips to
remove the 340~W dissipated in each calorimeter.
%
%\begin{figure}[ht]
%\centerline{\epsfxsize=8cm\epsfbox{fig1.eps}}   
%\caption{\label{fig:swdraw}}
%\end{figure}
%
The distribution of material upstream of the calorimeters is kept at a minimum
especially in the crucial region of the inner acceptance cut where it amounts to
0.25~$X_0$. In the middle of the acceptance this material increases to
about 2~$X_0$ due to cables
and support structures of the beam pipe. The effects of the degraded energy
resolution are important and are corrected for.

\begin{figure}[thb]
\centerline{\epsfxsize=6cm\epsfbox{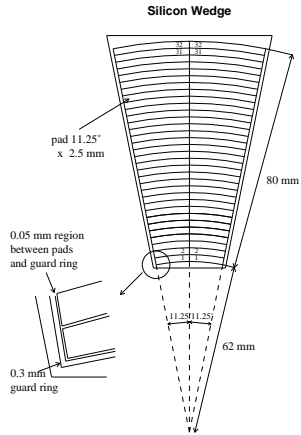}}   
\caption{Layout of the pad geometry of one wedge.
\label{fig:wedge}}
\end{figure}
Each detector wedge is a thick-film ceramic hybrid carrying a 64-pad silicon
wafer diode plus the readout electronics. The pad layout of the silicon diodes is
shown in Figure \ref{fig:wedge}. The pads are arranged in a $R-\phi$ geometry,
with a radial pitch of $2.5$~mm. Readout is done with 4 DC-coupled AMPLEX chips
(each one reading 16 channels in a given $\phi$ column). The diodes have an
average depletion voltage of 62~V and are operated at 80~V bias voltage.
The complete luminometer has in total $608$ wedges with a total of
$38,912$ readout channels.

For a typical LEP-I Bhabha electron with $E_e = 45$~GeV the charge deposited on
a single detector layer at shower maximum is $300-400$~mips 
($\approx 1.0-1.3$~pC) which is typically spread over a few pads.
The AMPLEX chip has a full scale limit of more than $1000$~mips for each pad,
thus providing
a sufficient dynamic range.
The equivalent noise for each channel remained at a level of 1500 to 2000
electrons for a typical detector capacitance of 20~pF, giving better than $10:1$
signal to noise for mips. 

The calibration has been studied with electrical pulses generated both on the
AMPLEX chips themselves and on the hybrids, as
well as with ionization signals generated in the Si using test beams and
laboratory sources.
The overall channel-to-channel uniformity in gain was 1~$\%$
but gain variations among the 16 channels of each AMPLEX were $\leq 0.25 \%$.
This allowed optimum resolution for trigger thresholds and eliminated the need
for a database of calibration constants for off-line energy reconstruction.
Cross talk among channels in each AMPLEX was at the level of $2
\%$/channel (coherent 30~$\%$/AMPLEX) and was subtracted.
Any residual gain variations depending on the channel position within each
AMPLEX were cancelled by inverting the channel radial ordering between the two
$\phi$ columns of each wedge.
 
The calorimeters were exposed to substantial radiation from occasional
catastrophic beam losses.
To limit this damage a protection system
monitored the bias currents and induced a fast beam dump if the absorbed
energy was greater than $3 \times 10^8$~GeV within 1~s.
The leakage current at $22^\circ$C was uniformly 1~nA/cm$^2$ when the detector was
installed in 1993. 
Radiation damage during eight years of operation at LEP
increased it to 12~nA/cm$^2$ on average, although at shower maximum the typical 
values are 5 times higher 
(the AMPLEX bias current limit is $\approx 200$~nA/pad).
From such increase of the leakage current we have estimated
an effective absorbed dose of about 4~Krad, 
or a total absorbed energy of $\approx 5 \times 10^{12}$~GeV, 
using measurements from J.~Lauber et al. \cite{bib:lauber}.
At the end of LEP running only $0.6 \%$ of the Si-W Luminometer
was not functional.

\section{Lateral shower profile}

\begin{figure}[h]
\centerline{\epsfxsize=7cm\epsfbox{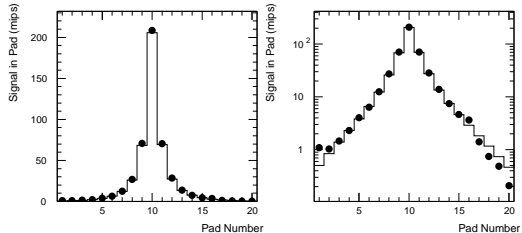}}   
\caption{Average radial shower profile at 6~$X_0$ for $E = 45.5$~GeV electrons
in linear ({\it left}) and logarithmic ({\it right}) scale.
\label{fig:shprof}}
\end{figure}

The lateral profile of electromagnetic showers in the dense medium of the Si-W
calorimeters is characterized by a sharp central peak (FWHM $<$~1 pad = 2.5~mm)
and broad tails extending to almost 10~pads, as shown in Fig.~\ref{fig:shprof}.
\begin{figure}[b]
\vspace*{-0.5cm}
\centerline{\epsfxsize=7cm\epsfbox{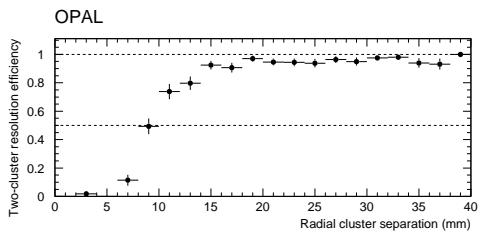}}   
\caption{Efficiency of reconstructing a secondary cluster as a function of the
radial separation with the primary one when they have equal azimuth.
\label{fig:eff2clus}}
\end{figure}

Peak finding is based on the second spatial derivative of the pad charge, so
that a sufficiently pronounced shoulder can be identified as a secondary
cluster. Radiative Bhabha events with one or more photons contained within the
acceptance can produce such configurations. The two cluster resolution
efficiency has been determined from such radiative data events 
with a well separated
secondary cluster with $E > 5$~GeV. The pad signals belonging to
the secondary cluster are rotated about the beam axis until they have the same
azimuth as the primary cluster and added to the signals actually observed on the local pads.
The standard reconstruction is then applied
and the separation efficiency as a function of the radial distance
between the two clusters is obtained, as shown in Fig.~\ref{fig:eff2clus}.
It is greater than $50 \%$ for cluster separation greater than 1.0~cm,
equivalent to 4 pad widths.
The overall inefficiency of primary cluster finding is less than $10^{-5}$.

\section{Position measurement}

The detector segmentation is very different in $R$
and $\phi$, owing to their different impact on the luminosity measurement.
Here we will be interested only in the precise radial position measurement.

The radial coordinate is first determined in each layer by interpolating a coordinate
within the pad displaying the maximum signal in that layer. 
Then all the good layer coordinates from 2~$X_0$ to 10~$X_0$ in depth are projected onto a
reference layer chosen at 7~$X_0$, and averaged there. The reference layer lies
near the average shower maximum to minimize systematic effects.  
The resolution of the layer coordinate varies strongly
across a pad, from about 300~$\mu$m at pad boundaries to 750~$\mu$m at pad
center. This variation is reflected even in the average $R$ coordinate, where a periodical
structure following the radial pitch is apparent. 
%(Fig.~\ref{fig:aveR}).
To remove such oscillation, as the last step, a smoothing algorithm is applied, 
subjected to boundary conditions at the pad boundaries.
%
%\begin{figure}[ht]
%\centerline{\epsfxsize=8cm\epsfbox{fig10a.eps}}   
%\caption{\label{fig:aveR}}
%\end{figure}
%

A key issue for the luminosity measurement is knowledge of the
absolute radial dimensions of the calorimeters. Very accurate positioning and
monitoring of detector wedges in each layer using microscopes and micro-manipulators
have achieved an RMS scatter of 1.3~$\mu$m of the radius of each wedge with respect to
the best-fit circle of each half-layer. Taking into account deviations of
each half-layer with respect to its ideal position in the calorimeter stack,
mechanical deformations, temperature effects and
measurement errors, the final precision on the absolute average radius is 
4.4~$\mu$m.

The final position resolution of the average smoothed radial coordinate has been
determined to be 130~$\mu$m at pad boundaries and 170~$\mu$m at pad centers,
from test beam measurements. The test beam used a 45~GeV electron beam
alternated with a 100~GeV muon beam. Alignment of the calorimeter with respect
to a high resolution Si-strip beam telescope was carried out with the muon beam.
Sensitivity of the Si-W electronics to mips was essential for this purpose.
The effect of upstream material was studied using a 0.84~$X_0$ plate which could be inserted in
front of the detector.

\begin{figure}[tb]
\centerline{\epsfxsize=7cm\epsfbox{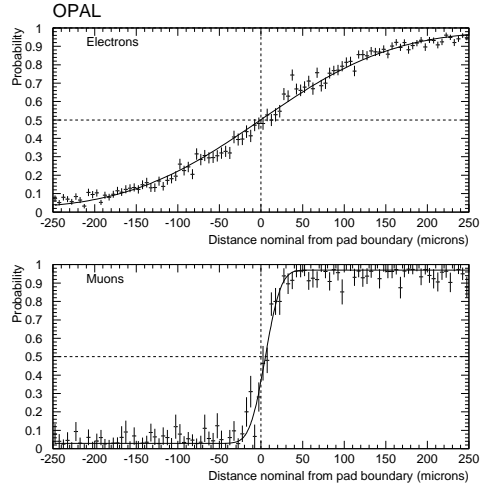}}   
\caption{Pad boundary images for test beam electrons and muons at 7~$X_0$.
\label{fig:pbimage}}
\end{figure}

\begin{figure}[tb]
\centerline{\epsfxsize=7cm\epsfbox{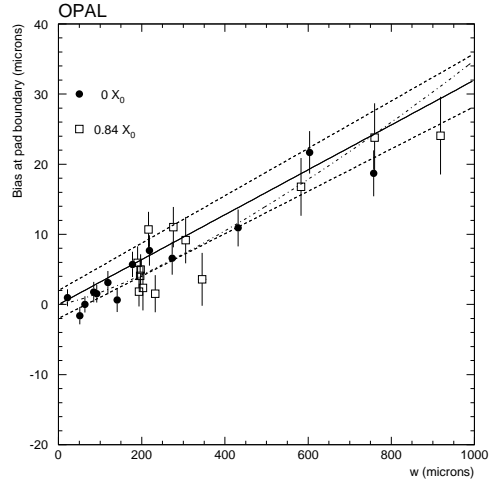}}   
\caption{The pad boundary bias as a function of the pad boundary transition
width $w$. The points refer to different depths in the bare calorimeter (solid
circles) or after an optional preshowering layer (open boxes).
\label{fig:pbbias}}
\end{figure}

The reconstruction method respects the symmetry condition that 
a shower which deposits equal energies on two adjacent pads 
in the reference layer at 7~$X_0$ has to be
reconstructed in the mean exactly at the boundary between the pads.
In reality due to the $R-\phi$ geometry of the pads, the true position of such
showers is at a smaller radius than the pad boundary. This is the so called {\em
pad boundary bias}, which depends on the lateral shower spread and has been
measured in the test beam.
As the radial position of the incoming particles is scanned across a radial pad boundary
in a single layer, the probability for observing the largest pad signal above or below
this boundary shifts rapidly, giving an image of the pad boundary as shown in Fig~\ref{fig:pbimage}.
The pad boundary images are modelled with an error function, where the gaussian
width $w$ is called the pad boundary transition width and $R_{off}$ is the radial
offset between the apparent and the nominal pad boundary. The difference in
$R_{off}$ obtained by changing from electron to muon beam is the measured pad
boundary bias, which is shown in Fig.~\ref{fig:pbbias} as a function of $w$. 
The reconstructed radial coordinate is sensitive to the distribution and type of
material in front of the detector as well as to the incidence angle of the
particles. The test beam configuration could not reproduce the exact features of
the OPAL running, so an indirect approach has been followed, called {\em
anchoring}. Details of the method are fully explained in the cited paper 
\cite{bib:lumi}.
The procedure is applied separately on individual data samples, each one
characterized by different beam parameters, 
and obtains net systematic corrections on the radius of the acceptance cuts. 
The inner acceptance cut is corrected by 5-10~$\mu$m  with an uncertainty 
of 3.5~$\mu$m, while the outer acceptance cut is corrected by 
10-20~$\mu$m with an uncertainty of 6~$\mu$m.
These radial corrections are then easily 
turned into acceptance corrections which are applied to data.

\begin{figure}[tb]
\centerline{\epsfxsize=7cm\epsfbox{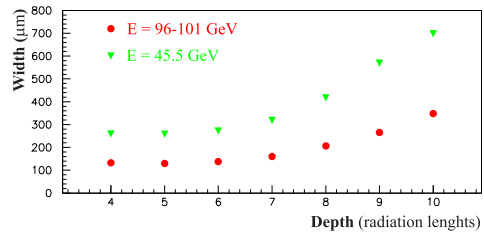}}   
\caption{Pad boundary transition width as a function of depth in the calorimeter
at a fixed radius for LEP-I and LEP-II energies. 
\label{fig:widthvsE}}
\end{figure}

We have also studied the energy dependence of the pad boundary transition width
using data from OPAL running, as there was no test beam data at LEP-II energies.
In Fig.~\ref{fig:widthvsE} $w$ is plotted as a function of depth into the
calorimeter for LEP-I and LEP-II Bhabha electrons. There is a sizeable shrinkage
of the shower core with increasing energy. As $w$ is related to the position
resolution near the pad boundaries, this indicates that the radial resolution
inherently improves at energies higher than LEP-I. 

\section{Energy measurement}

\begin{figure}[tb]
\centerline{\epsfxsize=9cm\epsfbox{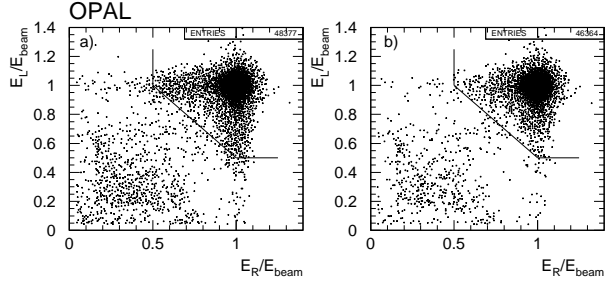}}   
\caption{Distribution of $E_L$ vs $E_R$ after all selection cuts except for the
energy cuts and before the acollinearity cut ({\it a}) or after it ({\it b}).
The lines show the cuts applied to the energies.
\label{fig:energies}}
\end{figure}

The distribution of the summed energy in the left and right calorimeters (after
all other cuts) is shown in Fig.~\ref{fig:energies} for a typical OPAL run.
The bulk of selected Bhabha events have back-to-back electrons and positrons
with energies close to the beam energy. 
The large accidental background is visible at small energies and 
is reduced to negligible levels by applying tight energy cuts,
which also eliminate a small fraction of real Bhabha events.
The visible radiative tails extending from the full energy spot originate from
events which have lost energy due to a single initial state photon emitted along
the beam axis. 
For these events transverse momentum conservation implies:
$E_R / E_L = R_L / R_R$.
%\begin{equation}
%\frac {E_R} {E_L} = \frac {R_L} {R_R}
%\end{equation}
A useful quantity to improve our understanding of the Bhabha events failing
the energy cut is the radius difference or acollinearity 
$\Delta R = R_R - R_L$. A cut at $\Delta R < 10$~mrad reduces both the
background and the impact of uncertainties in the low energy tail of the
detector response function, as can be seen from 
Fig.~\ref{fig:energies}/a-b.
The systematic error due to the energy measurement is reduced by almost a factor 3
with the $\Delta R$ cut. 
%and Fig.~\ref{fig:espec}. 
%Fig.~\ref{fig:espec} shows the normalized energy distribution in one of the
%calorimeters for selected Bhabha events, before and after application of the
%acollinearity cut, compared to the Monte Carlo prediction. In both cases
%excellent agreement is observed.
%\begin{figure}[tb]
%\centerline{\epsfxsize=4cm\epsfbox{fig31_cutai2.eps}}   
%%\begin{center}
%%\epsfig{file=fig31.eps,width=8cm
%%,bbllx=1pt,bblly=570pt,bburx=265pt,bbury=570pt,clip=
%%}
%%\end{center}
%\caption{\label{fig:espec}}
%\end{figure}
By cutting on the acollinearity one can also effectively limit or constrain 
the energy lost to initial state radiation. Therefore it is also useful to
provide clean samples of beam energy electrons for studying the energy response
of the calorimeters. Also samples with a selected lower energy can be 
isolated, though with lower statistics.
The energy resolution has stayed almost constant during all the LEP running.
At LEP-I ($E \approx 45$~GeV) $\Delta E /E = 3.8-4.5 \%$ (for right - left
calorimeter); at LEP-II ($E \leq 104$~GeV) $\Delta E /E = 5.0 \%$ (for both
right and left calorimeter). Differences between the two calorimeters as well as
from LEP-I and LEP-II are due to different amounts of preshowering material.

\section{Final error on luminosity}

The main experimental systematic errors on the OPAL luminosity measurement
at LEP-I \cite{bib:lumi} are summarized in table \ref{tab:errors}.
After all the effort on radial reconstruction, the dominant systematic error is
related to the energy measurement, mostly due to uncertainties in the tail
of the energy response function and the nonlinearity.
The final experimental systematic error successfully matches the desired level
of precision, well below $10^{-3}$, and even surpasses the present theoretical
precision of the calculated Bhabha cross
section, which is one of the most deeply studied QED processes.

\section{Conclusions}

The OPAL Si-W luminometer has reliably operated at LEP for 8 years
(1993-2000), with high efficiency and negligible losses of Si detectors and
readout electronics in a non-trivial background environment.  
Its performance can be summarized by these figures:
\begin{enumerate}
\item Energy resolution $\approx 4 \%$ almost constant from $E=45$~GeV to
\mbox{$E = 100$~GeV};
\item Good efficiency to resolve close lying clusters: $\epsilon \geq 50 \%$ for
\mbox{$\Delta R \geq 1.0$~cm};
\item Good S/N ratio for mips: $10/1$;
\item Position resolution on the radial coordinate of 130-170~$\mu$m with a
residual bias less than 7~$\mu$m.
\end{enumerate}
In particular the very small residual bias on the position of the acceptance 
cut was crucial to achieve the
extraordinary experimental systematic error of only $3.4 \times 10^{-4}$.

\begin{table}[tb]
\tbl{The most important systematic errors in the final luminosity measurement
for LEP-I.
\vspace*{1pt}}
{\footnotesize
%\tabcolsep7pt
%\begin{tabular}{@{}crrrr@{}}
\begin{tabular}{|c|r|}
\hline
{} &{} \\[-1.5ex]
Systematic errors &  $\times 10^{-4}$ \\[1ex]
\hline
{} &{} \\[-1.5ex]
Energy           & 1.8 \\[1ex]
Inner Anchor     & 1.4 \\[1ex]
Radial Metrology & 1.4 \\[1ex]
\hline
{} &{} \\[-1.5ex]
Total Experimental & 3.4 \\[1ex]
\hline
{} &{} \\[-1.5ex]
Total Theoretical & 5.4 \\[1ex]
\hline
\end{tabular}\label{tab:errors} }
\end{table}

\end{document}